\documentclass[12pt]{article}
\usepackage{epsfig}

\newcommand{\mysection}{\setcounter{equation}{0}\section}

\def\beq{\begin{equation}}
\def\eeq{\end{equation}}
\def\beqa{\begin{eqnarray}}
\def\eeqa{\end{eqnarray}}
 
\newlength{\dinwidth} \newlength{\dinmargin}
\begin{document}

\begin{center}
{\Large \bf Soft-gluon corrections for $tW$ production at N$^3$LO}
\end{center}
\vspace{2mm}
\begin{center}
{\large Nikolaos Kidonakis}\\
\vspace{2mm}
{\it Department of Physics, Kennesaw State University,\\
Kennesaw, GA 30144, USA}
\end{center}
 
\begin{abstract}
I present approximate results that include third-order soft-gluon corrections 
for the associated production of a single top quark with a $W$ boson. 
The calculation uses expressions from soft-gluon resummation at 
next-to-next-to-leading-logarithm (NNLL) accuracy.  
From the NNLL resummed cross section I derive approximate 
next-to-next-to-next-to-leading order (aN$^3$LO) 
cross sections for the process $bg \rightarrow tW^-$ at LHC and Tevatron 
energies. The aN$^3$LO top-quark transverse-momentum and rapidity distributions
in $tW$ production are also presented. 
\end{abstract}
 
\mysection{Introduction}
 
In the current state of particle physics and its exploration at collider energies, it is crucial to have a good theoretical understanding of top quark production cross sections and differential distributions. 
An important top production process at LHC energies is the associated production of a top quark with a $W$ boson, which proceeds via the partonic process $bg \rightarrow tW^-$ and is sensitive to the value 
of the $V_{tb}$ CKM matrix element as well as  possible new physics. 

Leading-order calculations and studies for $tW$ production and decays were presented in \cite{LY,HBB,SM,BBD} and with some additional corrections in \cite{TT,BB}. The complete next-to-leading order (NLO) corrections to $bg \rightarrow tW^-$ were calculated in Ref. \cite{ZhutW}. 
The NLO corrections are large and need to be taken into account in theoretical predictions. NLO corrections to $tW$ production including the decays of both the top quark and the $W$ boson were presented in Ref. \cite{JCFT}. Top-quark transverse-momentum, $p_T$, distributions in $tW$ production at NLO matched with parton showers appeared in Ref. \cite{Re}. Updated predictions for the total cross section at NLO have appeared in \cite{HATHOR}. For recent reviews see Refs. \cite{PF,Topc,NKtop,DL,AG,JWK}.

Soft-gluon corrections for $tW$ production were resummed at next-to-leading logarithm (NLL) accuracy in Ref. \cite{NKnll}. Fixed-order expansions of the NLL resummed cross section were also derived in \cite{NKnll} at NLO, next-to-NLO (NNLO), and next-to-NNLO (N$^3$LO). The resummation of soft-gluon contributions was extended via two-loop calculations to next-to-next-to-leading logarithm (NNLL) accuracy in Ref. \cite{NKtW}, where NLO and NNLO expansions of the NNLL resummed cross section were also provided. 

The soft-gluon corrections are an important component of the cross section and they constitute numerically the majority of the higher-order corrections, particularly near partonic threshold. The expansion of the resummed cross section provides approximate results at NLO and higher orders. It was shown in Ref. \cite{NKnll} that the approximate NLO cross section approximates very well the exact NLO result, and the higher-order soft-gluon corrections are significant. This is also in line with related results for single-top production in the $t$ and $s$ channels \cite{NKnll,NKtchsch, NKppn}, for top-antitop pair production \cite{NKppn, NKttbar}, and for $W$-boson production at large transverse momentum \cite{NKWZ}.

Approximate NNLO (aNNLO) cross sections were calculated for $tW$ production at NNLL accuracy in Refs. \cite{NKtW,NKppn}. Good agreement has been found between the theoretical predictions and recent experimental data from the LHC \cite{ATLAS7,CMS7,CMS8,ATLAS8,LHC8,ATLAS13}. The top-quark transverse momentum distribution in this process was presented at aNNLO in Ref. \cite{NKtWpT}. Some partial results for aNNLO top-quark rapidity distributions were given in \cite{NKconf1}.

In this paper we include third-order soft-gluon corrections from NNLL resummation for $tW$ production and provide approximate N$^3$LO (aN$^3$LO) total cross sections and top-quark transverse-momentum and rapidity distributions. In Section 2 we discuss soft-gluon resummation and present our analytical results for the soft-gluon corrections and their implementation. In Section 3 we present aN$^3$LO total cross sections for $tW$ production at LHC and Tevatron energies. In Section 4 we show aN$^3$LO top-quark $p_T$ and rapidity distributions in $tW$ production at LHC energies. We conclude in Section 5.

\mysection{Soft-gluon corrections for $bg \rightarrow tW^-$}

\begin{figure}
\begin{center}
\includegraphics[width=10cm]{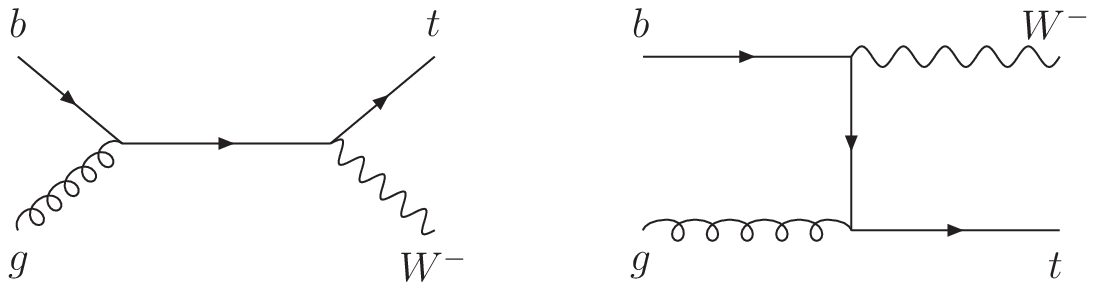}
\caption{Leading-order diagrams for $bg \rightarrow tW^-$.}
\label{tWborn}
\end{center}
\end{figure}

The leading-order diagrams for the process $bg \rightarrow tW^-$, involving a bottom quark and a gluon in the initial state, are shown in Fig. \ref{tWborn}.

We assign the momenta
\beq
b(p_1)\, + \, g\, (p_2) \rightarrow t(p_3)\, + W^-(p_4) \, , 
\eeq
and define $s=(p_1+p_2)^2$, $t=(p_1-p_3)^2$, $t_1=t-m_t^2$, $t_2=t-m_W^2$, 
$u=(p_2-p_3)^2$, $u_1=u-m_t^2$, and $u_2=u-m_W^2$, where $m_t$ is the 
top-quark mass and  $m_W$ is the $W$-boson mass. 
We also define the variable $s_4=s+t_1+u_2$ which 
measures distance from partonic threshold, where there is no energy for 
additional radiation, but the top quark and $W$-boson are not restricted 
to be produced at rest. 

The soft-gluon corrections appear in the perturbative expansion of the cross 
section as plus distributions of logarithms of $s_4$, specifically 
$[(\ln^k(s_4/m_t^2))/s_4]_+$, with $k$ taking values from 0 to $2n-1$ for the 
$n$th order corrections in the strong coupling, $\alpha_s$.

The plus distributions are defined by their integrals with functions $f$ 
(which in our case involve the soft-gluon coefficients that will be presented later in this section), as 
\beqa
\int_0^{s_4^{max}} ds_4 \, \left[\frac{\ln^k(s_4/m_t^2)}
{s_4}\right]_{+} f(s_4) &=&
\int_0^{s_4^{max}} ds_4 \frac{\ln^k(s_4/m_t^2)}{s_4} [f(s_4) - f(0)]
\nonumber \\ &&
{}+\frac{1}{k+1} \ln^{k+1}\left(\frac{s_4^{max}}{m_t^2}\right) f(0) \, .
\label{plus}
\eeqa

Resummation of soft-gluon contributions is a consequence of the factorization of the cross section into various functions that describe soft and collinear emission in the partonic process. We take moments of the partonic scattering cross section,  
${\hat \sigma}(N)=\int (ds_4/s) \;  e^{-N s_4/s} {\hat \sigma}(s_4)$, with $N$ the moment variable, and write a factorized expression in $n=4-\epsilon$ dimensions:
\beq
{\hat \sigma}^{tW}(N,\epsilon)= 
H^{tW} \left(\alpha_s(\mu)\right)\; S^{tW} 
\left(\frac{m_t}{N \mu},\alpha_s(\mu) \right)\;
\prod_{i=b,g} J_i\left (N,\mu,\epsilon \right) 
\label{factsigma}
\eeq 
where $\mu$ is the scale, $H^{tW}$ is the hard-scattering function, 
$S^{tW}$ is the soft-gluon function for non-collinear soft-gluon emission, 
and $J_i$ are jet functions which describe 
soft and collinear emission from the incoming partons.

The soft function $S^{tW}$ requires renormalization and its $N$-dependence 
can be resummed via renormalization group evolution \cite{NKnll,NKtW}. We have 
\beq
S_b^{tW}=Z^*_S \; S^{tW} \, Z_S
\eeq
where $S_b^{tW}$ is the unrenormalized quantity and $Z_S$ is a renormalization constant.
Thus  $S^{tW}$ satisfies the renormalization group equation
\beq
\left(\mu \frac{\partial}{\partial \mu}
+\beta(g_s, \epsilon)\frac{\partial}{\partial g_s}\right)\,S^{tW}
=-2 \, S^{tW} \, \Gamma_S^{tW}
\eeq
where $g_s^2=4\pi\alpha_s$; 
$\beta(g_s, \epsilon)=-g_s \epsilon/2 + \beta(g_s)$ 
with $\beta(g_s)$ the QCD beta function; and 
\beq
\Gamma_S^{tW}=\frac{dZ_S}{d\ln\mu} Z^{-1}_S
=\beta(g_s, \epsilon)  \frac{\partial Z_S}{\partial g_s} Z^{-1}_S
\eeq
is the soft anomalous dimension
that controls the evolution of the soft-gluon function $S^{tW}$.
We determine $\Gamma_S^{tW}$ from the coefficients of the ultraviolet poles of the relevant loop diagrams calculated in dimensional regularization \cite{NKnll,NKtW}.

With the two-loop soft-anomalous dimension we can resum the soft-gluon 
corrections at NNLL accuracy in moment space. 
For $tW^-$  production the resummed partonic cross section in moment space 
is given by \cite{NKtW} 
\beq
{\hat{\sigma}}^{res}(N) =   
\exp\left[\sum_{i=b,g} E_i(N_i)\right]
H^{tW}
\left(\alpha_s(\sqrt{s})\right) \;
S^{tW}\left(\alpha_s(\sqrt{s}/{\tilde N'})
\right) \; \exp \left[2\int_{\sqrt{s}}^{{\sqrt{s}}/{\tilde N'}} 
\frac{d\mu}{\mu}\; \Gamma_S^{tW}
\left(\alpha_s(\mu)\right)\right]  \, .
\label{resHS}
\eeq
The first exponent \cite{GS87,CT89} in Eq. (\ref{resHS}) 
resums soft and collinear corrections from the incoming $b$-quark and gluon
and is well known (see \cite{NKtW} for details).

We expand the soft anomalous dimension for $bg \rightarrow tW^-$ as
$\Gamma_S^{tW}=\sum_{n=1}^{\infty}(\alpha_s/\pi)^n \Gamma_S^{(n)}$.
The one-loop result is \cite{NKnll,NKtW}
\beq
\Gamma_S^{(1)}=C_F \left[\ln\left(\frac{-t_1}{m_t\sqrt{s}}\right)
-\frac{1}{2}\right] +\frac{C_A}{2} \ln\left(\frac{u_1}{t_1}\right)
\label{tW1l}
\eeq
with color factors $C_F=(N_c^2-1)/(2N_c)$ and $C_A=N_c$, 
where $N_c=3$ is the number of colors.

The two-loop result is \cite{NKtW}
\beq
\Gamma_S^{(2)}=\frac{K}{2} \Gamma_S^{(1)}
+C_F C_A \frac{(1-\zeta_3)}{4}
\nonumber 
\eeq
where $\Gamma_S^{(1)}$ is given in Eq. (\ref{tW1l}),  
$K=C_A(67/18-\zeta_2)-5n_f/9$ \cite{KR} with $n_f=5$ the number of 
light-quark flavors, $\zeta_2=\pi^2/6$, and $\zeta_3=1.2020569\cdots$.

We expand the moment-space resummed cross section, Eq. (\ref{resHS}), in the strong coupling, $\alpha_s$, invert to momentum space, and provide results through third order for the soft-gluon corrections. In other words, we use the resummed cross section in moment space as a generator of fixed-order results instead of deriving a resummed cross section in momentum space. The reason we do this is that fixed-order expansions do not require prescriptions to avoid divergences in the resummed expression, and they have predicted (see e.g. \cite{NKnll,NKtW,NKtchsch,NKttbar}) exact results very well for various top-quark processes. This is in contrast to resummed results that use prescriptions which have grossly underestimated the exact results. 

The analytical and numerical differences between our approach and prescription-based resummations have been described in detail in Ref. \cite{NK2000} where the theoretical reasons for the differences were explained. The differences arise due to unphysical subleading terms that are kept in minimal prescription approaches but not in the fixed-order expansions of our approach (for detailed discussions see Secs. IIIC and IV in \cite{NK2000}). These unphysical subleading terms do not appear in the exact results and their numerical impact is to underestimate the size of the exact corrections by a very wide margin. This is best illustrated in the related process of $t{\bar t}$ production where the exact NNLO corrections are large. These NNLO results were predicted extremely well by the aNNLO corrections in our formalism, far better than by any other resummation procedure. In fact the minimal prescription prediction for the NNLO corrections was smaller by an order of magnitude than our results (for additional discussions and comparisons see Refs. \cite{NKtop,NKppn,NKttbar,NKconf1,NKconf2}). For a review of various resummations for top-quark production see Ref. \cite{NKBP}. 

The differences between results from prescriptions and fixed-order expansions are very large, in fact much larger than higher-order terms at aN$^3$LO and beyond. The conclusion is that fixed-order approximations work best as one has better control of matching and subleading terms, and one does not need to worry about arbitrariness from prescription methods. We will also give some numerical discussion about these differences in the present context of $tW$ production in the next section; as we will see, the same conclusions will be drawn here as well, as expected, given the general nature of our considerations.

The NLO expansion of the resummed cross section for $tW$ production in momentum space is given in \cite{NKnll,NKtW} while the NNLO expansion is given at NLL 
in \cite{NKnll} and at NNLL in \cite{NKtW}.
The N$^3$LO soft-gluon corrections from the expansion of the resummed partonic cross section at NLL can be found in \cite{NKnll}. In general, the N$^3$LO soft-gluon corrections to the double-differential cross section can be calculated using the master formula in \cite{NNNLO}. 
Explicitly, we have 
\beqa
\frac{d^2{\hat{\sigma}}^{(3)}}{dt \, du}&=&F_{\rm LO} \frac{\alpha_s^3(\mu_R)}{\pi^3} \;  
\left\{(C_F+C_A)^3 \; \left[\frac{\ln^5(s_4/m_t^2)}{s_4}\right]_+ \right.
\nonumber \\ &&
{}+\frac{5}{2} \, (C_F+C_A)^2 \, \left[-(C_F+C_A)\ln\left(\frac{\mu_F^2}{s}\right)
+2 \, C_F \, \ln\left(\frac{t_1}{u_2}\right)
+C_F \, \ln\left(\frac{m_t^2}{s}\right)-C_F \right.
\nonumber \\ && \left. \left.
{}+C_A \, \ln\left(\frac{u_1}{t_1}\right)
-2 \, C_A \, \ln\left(\frac{-t_2}{m_t^2}\right) -\frac{11}{9} C_A
+\frac{2}{9} n_f\right] \;  
\left[\frac{\ln^4(s_4/m_t^2)}{s_4}\right]_+ +\cdots \right\}
\label{N3LOapprox}
\eeqa
where $\mu_F$ is the factorization scale, $\mu_R$ is the renormalization scale, and for brevity we have provided only the highest two powers of the logarithms. Here the leading-order factor is 
\beqa
\hspace{-8mm}F_{\rm LO}&=& 
\frac{\pi V_{tb}^2 \alpha_s \alpha}{12 m_W^2 \sin^2\theta_W s^2}
\left\{-\frac{1}{2 u_1^2} \left(2 m_W^2+m_t^2\right)
\left[u_2(s+3m_t^2-m_W^2)+t_1 (m_t^2- m_W^2)\right] \right.
\nonumber \\ && \hspace{-15mm} \left. 
{}+\frac{1}{u_1 s}\left[2t_1(m_t^2-m_W^2)m_W^2+u_2 (t_1+u_1) m_t^2
+s m_t^2 (2 m_W^2+m_t^2)\right]
-\frac{u_1}{2s} \left[2 m_W^2+m_t^2\right] \right\}
\eeqa
where $\alpha=e^2/(4\pi)$ and $\theta_W$ is the weak mixing angle.
We note that with NNLL accuracy all soft-gluon logarithm terms can be fully determined at NNLO, but only the terms with the four highest powers of the 
logarithms can be fully determined at N$^3$LO.

The result for the third-order soft-gluon corrections can be written compactly as 
\beq 
\frac{d^2{\hat \sigma}^{(3)}}{dt \, du}=F_{\rm LO} \frac{\alpha_s^3}{\pi^3} 
\sum_{k=0}^5 C_k^{(3)} \left[\frac{\ln^k(s_4/m_t^2)}{s_4}\right]_+
\label{3corr}
\eeq
where the $C_k^{(3)}$ denote the coefficients of the logarithms: for example,  
$C_5^{(3)}=(C_F+C_A)^3$.

\mysection{aN$^3$LO total cross sections for $tW$ production}

We consider proton-proton (or proton-antiproton) collisions with momenta $p_A+p_B \rightarrow p_3+p_4$. We define the hadronic kinematical variables  
$S=(p_A+p_B)^2$, $T=(p_A-p_3)^2$, $T_1=T-m_t^2$, $T_2=T-m_W^2$, 
$U=(p_B-p_3)^2$, and $U_1=U-m_t^2$. They are related to the partonic variables defined earlier via the relations $p_1=x_1 p_A$ and $p_2=x_2 p_B$, where $x_1$ and $x_2$ are the fractions of the momentum carried by the partons in hadrons $A$ and $B$, respectively.

The hadronic total cross section can be written as 
\beqa
\sigma_{tW}&=& \int_{T^{min}}^{T^{max}} dT \int_{U^{min}}^{U^{max}} dU
\int_{x_2^{min}}^1 dx_2 \int_0^{s_4^{max}} ds_4 \, 
\frac{x_1 \, x_2}{x_2 S+T_1} \,
\phi(x_1) \, \phi(x_2) \, 
\frac{d^2{\hat\sigma}}{dt \, du}
\nonumber \\ &&
\label{totalcs}
\eeqa
where the $\phi$ denote the parton distribution functions (pdf);   
$x_1=(s_4-m_t^2+m_W^2-x_2U_1)/(x_2 S+T_1)$; 
$T^{^{max}_{min}}=-(1/2)(S-m_t^2-m_W^2) \pm 
(1/2) [(S-m_t^2-m_W^2)^2-4m_t^2m_W^2]^{1/2}$; 
$U^{max}=m_t^2+S m_t^2/T_1$ and $U^{min}=-S-T_1+m_W^2$;
$x_2^{min}=-T_2/(S+U_1)$; 
and $s_4^{max}=x_2(S+U_1)+T_2$.

In particular, using Eq. (\ref{3corr}) and the properties of plus distributions, Eq. (\ref{plus}), the aN$^3$LO corrections to the total cross section, Eq. (\ref{totalcs}), can be written as 
\beqa
\sigma^{(3)}_{tW}&=& \frac{\alpha_s^3}{\pi^3} 
\int_{T^{min}}^{T^{max}} dT \int_{U^{min}}^{U^{max}} dU
\int_{x_2^{min}}^1 dx_2 \, \phi(x_2) \frac{x_2}{x_2 S+T_1}
\nonumber \\ && \hspace{-20mm}
\times \sum_{k=0}^5 \left\{\int_0^{s_4^{max}} ds_4 
\frac{1}{s_4} \ln^k\left(\frac{s_4}{m_t^2}\right) 
\left[F_{\rm LO} \, C_k^{(3)} \, x_1 \, \phi(x_1)
-F_{\rm LO}^{\rm el} \, C_k^{(3) \rm el} \, x_1^{\rm el} \, \phi\left(x_1^{\rm el}
\right)\right] \right.
\nonumber \\ && \hspace{-17mm} \left. 
{}+\frac{1}{k+1} \ln^{k+1}\left(\frac{s_4^{max}}{m_t^2}\right) 
F_{\rm LO}^{\rm el} \, C_k^{(3) \rm el} \, x_1^{\rm el} \, \phi\left(x_1^{\rm el}\right)  \right\} \, .
\eeqa
where $x_1^{\rm el}$, $F_{\rm LO}^{\rm el}$, and $C_k^{(3)\, \rm el}$ denote the 
elastic variables, calculated with $s_4=0$.

\begin{table}[htb]
\begin{center}
\begin{tabular}{|c|c|c|c|c|} \hline
\multicolumn{5}{|c|}{aN$^3$LO $tW^-$ cross section (pb)} \\ \hline
$m_t$ (GeV) & LHC 7 TeV & LHC 8 TeV & LHC 13 TeV & LHC 14 TeV \\ \hline
172.5 & 8.5 $\pm$ 0.2 $\pm$ 0.3 & 12.0 $\pm$ 0.3 $\pm$ 0.4 & 38.1 $\pm$ 0.9 $\pm$ 0.9 & 44.8  $\pm$ 1.0 $\pm$ 1.0 \\ \hline
173.3 & 8.3 $\pm$ 0.2 $\pm$ 0.3 & 11.8 $\pm$ 0.3 $\pm$ 0.4 & 37.6 $\pm$ 0.9 $\pm$ 0.9 & 44.3  $\pm$ 1.0 $\pm$ 1.0 \\ \hline 
\end{tabular}
\caption[]{The aN$^3$LO $bg \rightarrow tW^-$ production cross section in pb 
in $pp$ collisions at the LHC with $\sqrt{S}=7$, 8, 13, and 14 TeV, 
using the MMHT2014 NNLO pdf \cite{MMHT2014}.}
\label{table1}
\end{center}
\end{table}

\subsection{Cross sections at LHC energies}

We now use these analytical expressions to calculate aN$^3$LO cross sections 
for $tW$ production via the process $bg \rightarrow tW^-$ at the LHC and the 
Tevatron.

We begin with results for the total cross section using MMHT2014 NNLO pdf 
\cite{MMHT2014}. In Table 1 we provide numerical values for the 
aN$^3$LO $tW^-$ cross section at the LHC for energies of 7, 8, 13, and 14 TeV,  
and two different values of top quark mass, $m_t=172.5$ and 173.3 GeV. 
The central value is calculated with $\mu=m_t$.
The first uncertainty is from scale variation between $m_t/2$ and $2m_t$ 
and the second is from the MMHT2014 NNLO pdf at 68\% C.L.
As is already known from \cite{NKtW}, the NNLO soft-gluon corrections increase 
the NLO cross section by a sizable amount, of the order of 10\%. 
The aN$^3$LO corrections further increase the aNNLO cross section by $\sim 4$\%.
We note that the cross section for 
${\bar b} g \rightarrow {\bar t} W^+$ is identical.

At the Tevatron with 1.96 TeV energy the cross section 
for $bg \rightarrow tW^-$ is very small and 
we find the value 0.100 $\pm$ 0.004 $\pm$ 0.010 pb for $m_t=173.3$ GeV. We note that $tW$ production has not been observed at the Tevatron due to the small size of the cross section.

\begin{figure}
\begin{center}
\includegraphics[width=10cm]{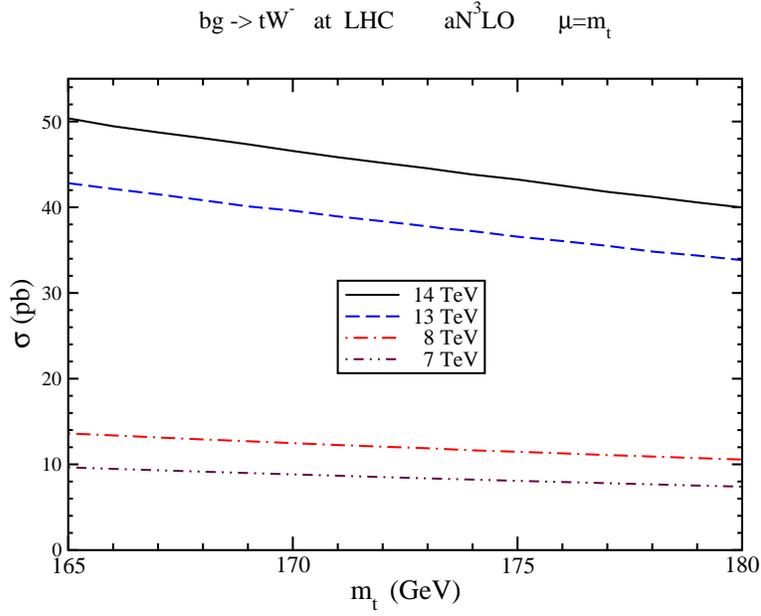}
\caption{The  aN$^3$LO cross section for $tW^-$ production 
at the LHC with $\sqrt{S}=7$, 8, 13, and 14 TeV.}
\label{LHCtW}
\end{center}
\end{figure}

In Fig. \ref{LHCtW} we plot the aN$^3$LO cross section
for $bg \rightarrow tW^-$ versus top quark mass 
for LHC energies of 7, 8, 13, and 14 TeV. The dependence on the top-quark mass 
is relatively mild given the small current uncertainties on the mass. Even in 
the wide mass range from 165 to 180 GeV shown in the figure, the variation of
the cross section is of order 20\%.

\begin{figure}
\begin{center}
\includegraphics[width=10cm]{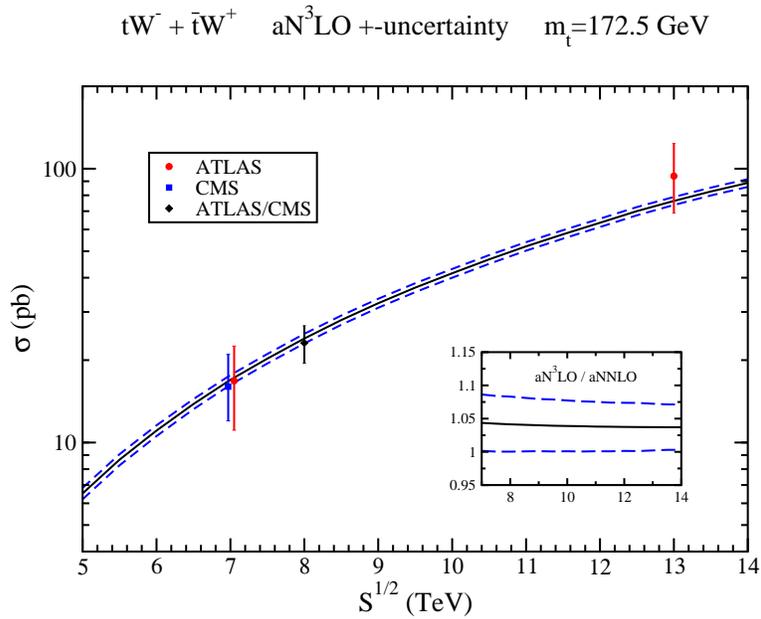}
\caption{The aN$^3$LO sum total cross section for $tW$ production (sum of $tW^-$ and ${\bar t}W^+$) as a function of the LHC energy. The inset plot displays the ratio of the aN$^3$LO cross section, with theoretical uncertainty, to the central aNNLO result at 13 TeV.}
\label{LHCStW}
\end{center}
\end{figure}

In Fig. \ref{LHCStW} we plot the aN$^3$LO cross section sum total 
for $tW^-$ and ${\bar t} W^+$ production, i.e. the sum of the 
$bg \rightarrow tW^-$ and ${\bar b} g \rightarrow {\bar t}W^+$ cross sections, 
as a function of $\sqrt{S}$. The central result is shown together with maximum and minimum values arising from the total theoretical uncertainty.
We compare with LHC data at 7 TeV from ATLAS \cite{ATLAS7} and CMS \cite{CMS7}, at 8 TeV from an ATLAS/CMS combination \cite{LHC8}, and at 13 TeV from ATLAS \cite{ATLAS13}. We observe excellent agreement of the theoretical curves with all LHC data.

The inset plot in Fig. \ref{LHCStW} displays the ratio of the aN$^3$LO cross section (central $\mu=m_t$ value, and maximum/minimum values from the uncertainties) to the central aNNLO cross section. The aN$^3$LO/aNNLO ratio for the central value is around 4\% and, as expected, it is somewhat higher at lower LHC energies where the threshold region is more important.

We note that results using other pdf sets are similar. In past work \cite{NKtW} the older MSTW2008 NNLO pdf \cite{MSTW2008} were used. The MMHT2014 pdf slightly increase the cross section relative to MSTW2008 as also noted in Ref. \cite{NKtWpT}. If one uses the CT14 NNLO pdf \cite{CT14} the central result is essentially the same to that in the table but the pdf uncertainties are larger, e.g. $37.6 \pm 0.9 \pm 1.7$ pb at 13 TeV energy for $m_t=173.3$ GeV. The CT14 pdf uncertainties are similar to the 90\% C.L. MSTW2008 pdf uncertainties, while the MMHT2014 pdf uncertainties at 68\% C.L. are similar to the ones from the MSTW2008 pdf at 68\% C.L. 

We also note that ideally one should use N$^3$LO pdf for the aN$^3$LO predictions. However, since N$^3$LO pdf are not available, the best choice at present is to use NNLO pdf. The change is already very small between aN$^3$LO results using NLO pdf and NNLO pdf (0.1 pb difference at 13 TeV energy), and thus we would expect a very small difference from N$^3$LO pdf if such pdf were available.

\subsection{Effects of subleading terms}

As noted before, at NNLO the coefficients of all the powers of the soft logarithms can be fully determined; but at N$^3$LO  only the coefficients of the four highest powers of the logarithms can be fully determined. However, we can still calculate the dominant terms in the coefficients of the lowest two powers of the logarithms at N$^3$LO; these terms involve constants from the Riemann $\zeta$ function ($\zeta_2$, $\zeta_3$, $\zeta_4$ and $\zeta_5$ at aN$^3$LO) that arise from the inversion from moment space to momentum space (see Ref. \cite{NK2000} for extended discussions on the structure of these $\zeta$ terms). In fact we determine such terms exactly, i.e. as they would appear in a full calculation.

The fact that such $\zeta$ terms are dominant in the coefficients of subleading logarithms can already be clearly seen at NNLO. For specificity let us consider $tW$ production at 13 TeV LHC energy, and for convenience let us use the notation $P_k \equiv [\ln^k(s_4/m_t^2)/s_4]_+$. We now study the progression of the contributions to the cross section as successive logarithmic terms are included. At aNNLO the $P_3$ terms (which are the leading logarithm terms) are 9.0 pb, while the sum of the $P_3$ and the $P_2$ terms is 7.3 pb. The sum of the $P_3$, $P_2$, and $P_1$ terms is 4.2 pb. Finally, the sum of the $P_3$, $P_2$, $P_1$, and $P_0$ terms (which is the complete aNNLO correction) is 2.6 pb. Now, if instead we sum the $P_3$ and $P_2$ terms and we add to them only the $\zeta$ contributions in the $P_1$ and $P_0$ terms, then we find 2.5 pb. The difference between 2.6 and 2.5 pb is negligible. This shows that the lower powers of the logarithms are dominated by $\zeta$ terms and it would be a very good approximation to use them if the complete $P_1$ and $P_0$ terms were not known. We find similar results at 8 TeV LHC energy: 0.88 pb for the complete aNNLO correction vs 0.84 pb for the one with incomplete $P_1$ and $P_0$ terms. It is also worth noting that very similar behavior is found in various other processes such as $t{\bar t}$ \cite{NKttbar,NK2000} production and $b{\bar b}\rightarrow H$ \cite{NKbbH}, so this seems to be a robust and widespread feature of soft-gluon corrections.  

Similarly at N$^3$LO, one can see that the subleading terms are dominated numerically by $\zeta$ constants. At aN$^3$LO the $P_5$ terms (which are the leading logarithm terms) are 14.5 pb at 13 LHC TeV energy. The sum of the $P_5$ and $P_4$ terms is 13.4 pb, and the sum of the $P_5$, $P_4$, and $P_3$ terms is 6.1 pb. Also the sum of the $P_5$, $P_4$, $P_3$, and $P_2$ terms is 1.6 pb; adding to that sum the $\zeta$ constants in the $P_1$ and $P_0$ terms we find 1.4 pb, which is our final result for the aN$^3$LO soft-gluon corrections. Based on the previous considerations at aNNLO, we expect that although at aN$^3$LO the $P_1$ and $P_0$ terms are incomplete, we have included in them the dominant, i.e. the $\zeta$, contributions. This expectation is further strengthened by the numerical dominance of the $\zeta$ terms in the coefficients where the full result is known, and it is further corroborated by analogous results for other processes including $t{\bar t}$ production, as well as the process $b{\bar b} \rightarrow H$ for which the full aN$^3$LO correction is known exactly \cite{NKbbH}. In any case, there is a very small difference between the results including or not including the subleading $P_1$ and $P_0$ terms (1.4 vs 1.6 pb).

As discussed in the previous section, for our numerical results we do not use a resummation because that would require a prescription and involve unphysical subleading terms (terms involving the Euler constant $\gamma_E$) in addition to the exact and physical subleading terms discussed in the preceding paragraphs. The validity and robustness of our fixed-order approach has been amply confirmed by its success in predicting the exact results at NNLO for $t{\bar t}$ production as well as postdicting the NLO results for $t{\bar t}$ and single-top production as well as other processes, as has already been discussed plenty of times before (see e.g. Refs. \cite{NKttbar,NKconf1,NKconf2}). Since $tW$ production is similar to the $t {\bar t}$ process in mass scale and in the size of both the aNNLO and aN$^3$LO corrections, it is instructive to again discuss the differences between fixed-order and resummed approaches here and show once again that it is best to use the fixed-order expansions. Since corrections beyond aN$^3$LO are negligible, the numerical differences between the approaches are due to unphysical subleading terms in the minimal prescription at NLO and NNLO, and to a lesser extent at aN$^3$LO.

We begin by noting that our predictions for soft-gluon corrections at NLO approximate very well the exact NLO results, as already discussed before in \cite{NKnll}. However, if one includes unphysical subleading terms in the inversion from moment space as used in the minimal prescription (but not in our approach, see Sec. IIIC of Ref. \cite{NK2000}), then the NLO corrections are reduced by a factor of two or more at all LHC energies as well as at Tevatron energy. For example, at 13 TeV LHC energy the results with unphysical subleading terms are 40\% of the exact value. This already indicates that the prescription approach is very problematic. 

At NNLO, the soft-gluon corrections are again reduced significantly at all energies if unphysical subleading terms are included, the difference being of the order of 30\%. At aN$^3$LO we again find similar differences between the two approaches at all energies. The NLO and NNLO soft-gluon terms dominate the corrections, with a smaller contribution from aN$^3$LO and negligible contributions beyond that; therefore, the difference between the predictions for the higher-order corrections is mostly due to that at NLO and NNLO, which we find to be very big. Thus, the total higher-order corrections (i.e. beyond leading order) are a factor of two smaller in the minimal resummed result than in our fixed-order expansion. We again note that our results for the NLO and NNLO soft-gluon corrections are complete, and thus include all soft-gluon terms. Since corrections beyond N$^3$LO are negligible in both approaches, the minimal resummed results miss the numerical importance of the soft gluons because of the vast underprediction of the true size of the NLO, NNLO, and aN$^3$LO corrections, in stark contrast to our results.

\subsection{$tW$ and $t{\bar t}$ interference}

A topic that has been previously discussed at great length in the literature is the intereference between $tW$ production and top-pair production \cite{TT,BB,ZhutW,JCFT,Re,HATHOR,PF,FLMWW,CKMP,JLNOP}. Starting at NLO, there are diagrams contributing to $tW$ production that can be thought of as top-pair production with decay of the antitop. Various procedures have been suggested to deal with this interference, including introduction of a cut on the invariant mass of the $W{\bar b}$ system to avoid resonance of the antitop propagator, diagram subtraction (implementing a gauge-invariant subtraction term in the cross section), diagram removal (excluding all NLO diagrams that are doubly resonant), etc. Approaches have been proposed in both the five-flavor and four-flavor schemes and implemented in various Monte Carlo generators.

Experimentally, appropriate selection cuts are made and discriminants are constructed to separate  the $tW$ signal from  the top-pair background \cite{ATLAS7,CMS7,CMS8,ATLAS8,LHC8,ATLAS13}. Diagram removal and diagram subtraction schemes have been used by both ATLAS and CMS to generate events for simulation samples, and consistency has been found between the two approaches.

The interference problem does not directly concern our calculations here. Our treatment of soft-gluon corrections and their resummation is based on LO, i.e. $2 \rightarrow 2$, kinematics. In other words we consider soft-gluon emission from the partons in the diagram of Fig. \ref{tWborn}. Therefore no diagram overlap exists for the soft-gluon corrections between the diagram for $tW$ production, i.e. $bg \rightarrow tW^-$, and the diagrams for $t{\bar t}$ production, i.e $q{\bar q} \rightarrow t {\bar t}$ and $gg \rightarrow t{\bar t}$. The higher-order soft-gluon corrections in our work are important as they significantly enhance the $tW$ cross section.

\mysection{aN$^3$LO top-quark $p_T$ and rapidity distributions}

We continue with top-quark differential distributions in $tW$ production.
We use MMHT2014 NNLO pdf \cite{MMHT2014} in our numerical results.

The top-quark transverse momentum, $p_T$, distribution can be written as 
\beqa
\frac{d\sigma}{dp_T}&=&
2 \, p_T \int_{Y^{min}}^{Y^{max}} dY \int_{x_2^{min}}^1 dx_2 
\int_0^{s_4^{max}} ds_4 \, 
\frac{x_1 \, x_2 \, S}{x_2 S+T_1} \,
\phi(x_1) \, \phi(x_2) \, 
\frac{d^2{\hat\sigma}}{dt \, du}
\nonumber \\ &&
\label{dpt}
\eeqa
where $T_1=-\sqrt{S} \, (m_t^2+p_T^2)^{1/2} \, e^{-Y}$,  
$U_1=-\sqrt{S} \, (m_t^2+p_T^2)^{1/2} \, e^{Y}$, and 
$Y^{^{max}_{min}}=\pm (1/2) \ln[(1+\beta_T)/(1-\beta_T)]$ with 
$\beta_T=[1-4(m_t^2+p_T^2)S/(S+m_t^2-m_W^2)^2]^{1/2}$.
We note that the total cross section can be obtained by integrating 
the $p_T$ distribution, $d\sigma/dp_T$, over $p_T$ from 0 to  
$p_T^{max}=[(S-m_t^2-m_W^2)^2-4m_t^2m_W^2]^{1/2}/(2\sqrt{S})$, 
and we have checked for consistency that we get the same numerical results 
as in the previous section.

\begin{figure}
\begin{center}
\includegraphics[width=11cm]{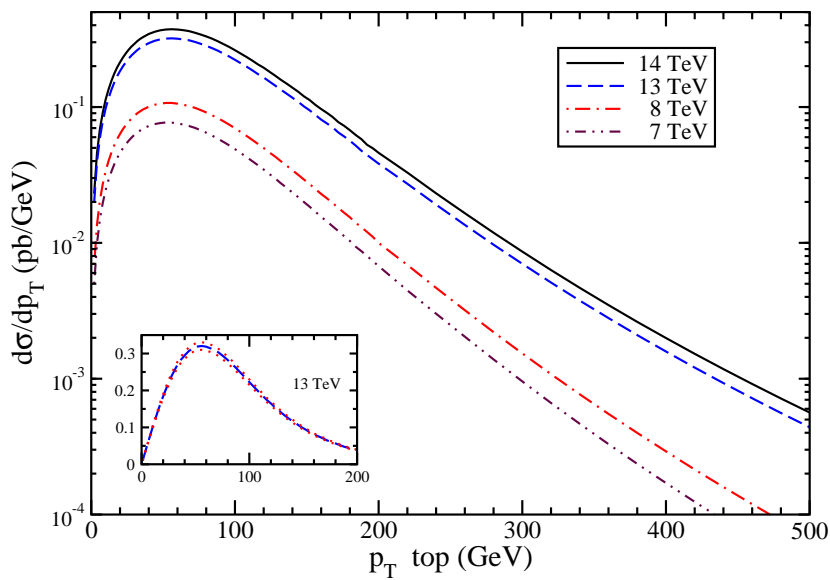}
\caption{The aN$^3$LO top-quark $p_T$ distribution in $tW^-$ production 
at the LHC with $\sqrt{S}=7$, 8, 13, and 14 TeV. The inset plot shows the distribution at 13 TeV with theoretical uncertainty.}
\label{pttoptW}
\end{center}
\end{figure}

In Fig. \ref{pttoptW} we plot the aN$^3$LO top $p_T$ distribution,
$d\sigma/dp_T$, in $bg \rightarrow tW^-$ production at LHC energies of 
7, 8, 13, and 14 TeV. The results vary over three orders of magnitude from 
the maximum values, at $p_T$ between 50 and 60 GeV, to the value at a $p_T$ 
of 500 GeV. In the inset plot we show the aN$^3$LO top $p_T$ distribution at 
13 TeV energy (central value at $\mu=m_t$, and maximum/minimum values from 
the total theoretical uncertainty) in a linear plot.

The top-quark rapidity, $Y$, distribution can be written as 
\beqa
\frac{d\sigma}{dY}&=&
\int_0^{p_T^{max}} 2 \, p_T \, dp_T \int_{x_2^{min}}^1 dx_2
\int_0^{s_4^{max}} ds_4 \, 
\frac{x_1 \, x_2 \, S}{x_2 S+T_1} \,
\phi(x_1) \, \phi(x_2) \, 
\frac{d^2{\hat\sigma}}{dt \, du}
\nonumber \\ &&
\label{dy}
\eeqa
where $p_T^{max}=((S+m_t^2-m_W^2)^2/(4S\cosh^2Y)-m_t^2)^{1/2}$.
We note that the total cross section can also be obtained by integrating 
the rapidity distribution, $d\sigma/dY$, over rapidity with limits 
$Y^{^{max}_{min}}=\pm (1/2) \ln((1+\beta)/(1-\beta))$
where $\beta=\sqrt{1-4m_t^2/S}$, and again we have checked for consistency 
that we get the same numerical results as in the previous section.

\begin{figure}
\begin{center}
\includegraphics[width=11cm]{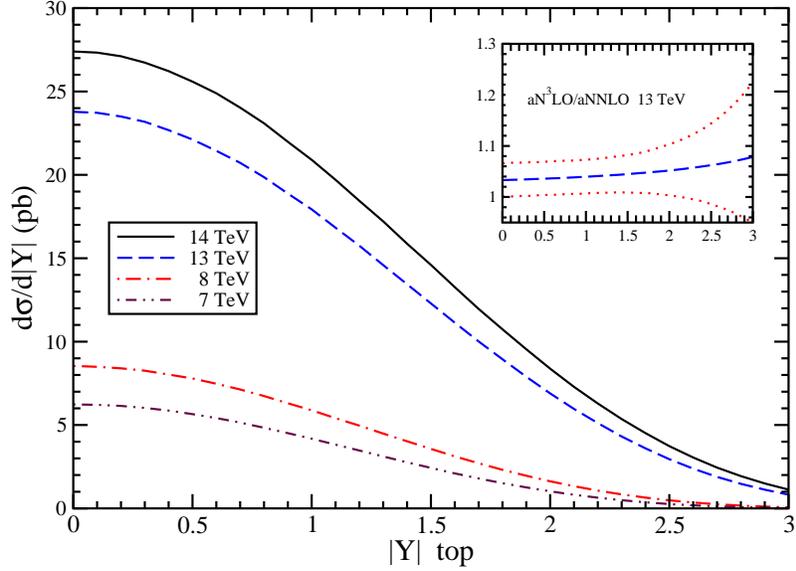}
\caption{The aN$^3$LO top-quark rapidity distribution in $tW^-$ production  
at the LHC with $\sqrt{S}=7$, 8, 13, and 14 TeV. The inset plot displays the ratio of the aN$^3$LO result, with theoretical uncertainty, to the central aNNLO result at 13 TeV.}
\label{yabstoptW}
\end{center}
\end{figure}

In Fig. \ref{yabstoptW} we plot the aN$^3$LO top rapidity distribution, 
$d\sigma/d|Y|$, in $bg \rightarrow tW^-$ production at 
energies of 7, 8, 13, and 14 TeV.
In the inset plot we show the ratio of the aN$^3$LO result (central $\mu=m_t$ value, and maximum/minimum values from total theoretical uncertainty) to the central aNNLO result at 13 TeV energy. As expected, the central ratio increases with $|Y|$, reaching almost 
an 8\% increase at $|Y|=3$. We also observe that the theoretical uncertainty 
increases at large $|Y|$.

\begin{figure}
\begin{center}
\includegraphics[width=11cm]{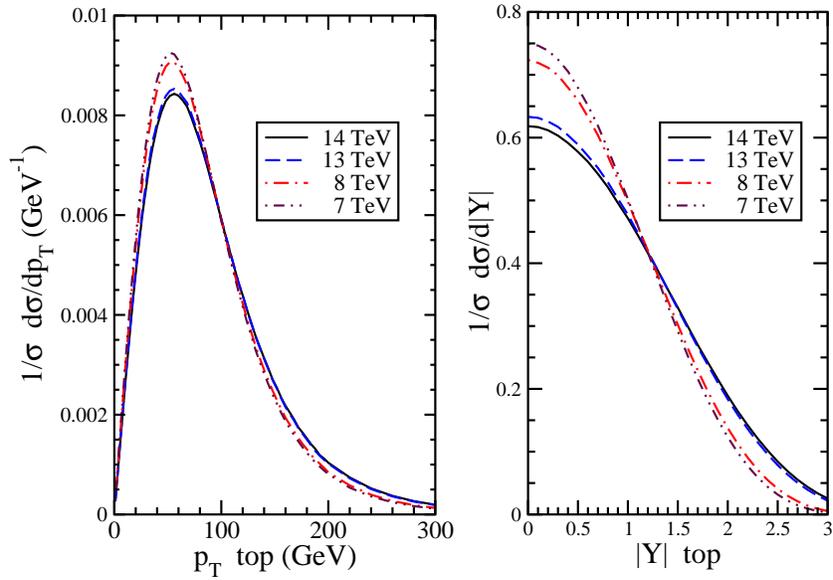}
\caption{The aN$^3$LO normalized top-quark $p_T$ (left) and rapidity (right) 
distributions in $tW^-$ production at LHC energies.}
\label{ptyabsnormtoptW}
\end{center}
\end{figure}

In Fig. \ref{ptyabsnormtoptW} we plot the normalized aN$^3$LO top $p_T$ 
distribution $(1/\sigma) d\sigma/dp_T$ (left), and the normalized aN$^3$LO top 
rapidity distribution $(1/\sigma) d\sigma/d|Y|$ (right), 
in $bg \rightarrow tW^-$ production at LHC energies. Normalized distributions minimize the effects of choices of different pdf sets and are thus often used in comparisons of data with theoretical predictions. At larger LHC energies the curves get higher than those for smaller energies at higher $p_T$ and $|Y|$ values, as expected.

\mysection{Conclusion}

The cross sections for the associated production of a single top quark with 
a $W$ boson, via $bg \rightarrow tW^-$, receive large contributions from 
soft gluon corrections. 
These soft-gluon contributions have been resummed to NNLL accuracy  
via two-loop soft anomalous dimensions. 
From the NNLL resummed formula approximate N$^3$LO double-differential
cross sections were derived. 
Numerical predictions were provided for the total cross section for 
$tW$ production at LHC and Tevatron energies. 
The aN$^3$LO corrections enhance the 
aNNLO cross section for $tW^-$ production at the LHC by $\sim$4\%.
The top-quark transverse-momentum and rapidity distributions were also 
presented at aN$^3$LO for LHC energies.

\mysection*{Acknowledgements}
This material is based upon work supported by the National Science Foundation under Grant No. PHY 1519606.

\end{document}